\documentclass[oneside,article,12pt]{memoir}

\usepackage{lineno,hyperref,graphicx,siunitx,textcomp,booktabs,miller,amsmath,adjustbox,gensymb,xcolor,sidecap,multicol,multirow}
\usepackage[version=4]{mhchem}
\usepackage{caption}
\setsecnumdepth{subsection}
\DoubleSpacing

\DeclareSIUnit{\atpercent}{at.\%}
\DeclareSIUnit{\at}{at.}

\newcommand\hl[1]{%
	\bgroup
	\hskip0pt\color{red!80!black}%
	#1%
	\egroup
}

\begin{document}

\begin{frontmatter}
 
\title{Explaining the Entropy Forming Ability with the atomic size mismatch}
\date{\vspace{-5ex}}
\maketitle



{\setlength{\parindent}{0pt}

Andreas Kretschmer*, Paul Heinz Mayrhofer

\vspace{5mm}

andreas.kretschmer@tuwien.ac.at, paul.mayrhofer@tuwien.ac.at

\vspace{5mm}

* Corresponding Author

TU Wien, Institute of Materials Science and Technology E308

Gumpendorferstrasse 7, 1060 Vienna, Austria

+43 1 58801-308108
}

\newpage

\DoubleSpacing

\begin{abstract}
	To quickly screen for single-phased multi-principal-element materials, a so-called entropy forming ability (EFA) parameter is sometimes used as a descriptor. The higher the EFA, the higher is the propensity to form a single-phase structure, which is stabilized against separation up to a certain threshold by the configurational entropy. We have investigated this EFA descriptor with atomic relaxations in special-quasi-random structures and discovered that the EFA correlates inversely with the lattice distortion. Large atomic size differences lead to multi-phase compounds, and little size differences to single-phase compounds. Instead of configurational entropy, we therefore demonstrate the applicability of the Hume-Rothery rules to phase stability of solid solutions even in compositionally complex ceramics.
	
\end{abstract}

\end{frontmatter}

\begin{figure}
	\makebox[\textwidth][c]{
		\includegraphics[width=19cm]{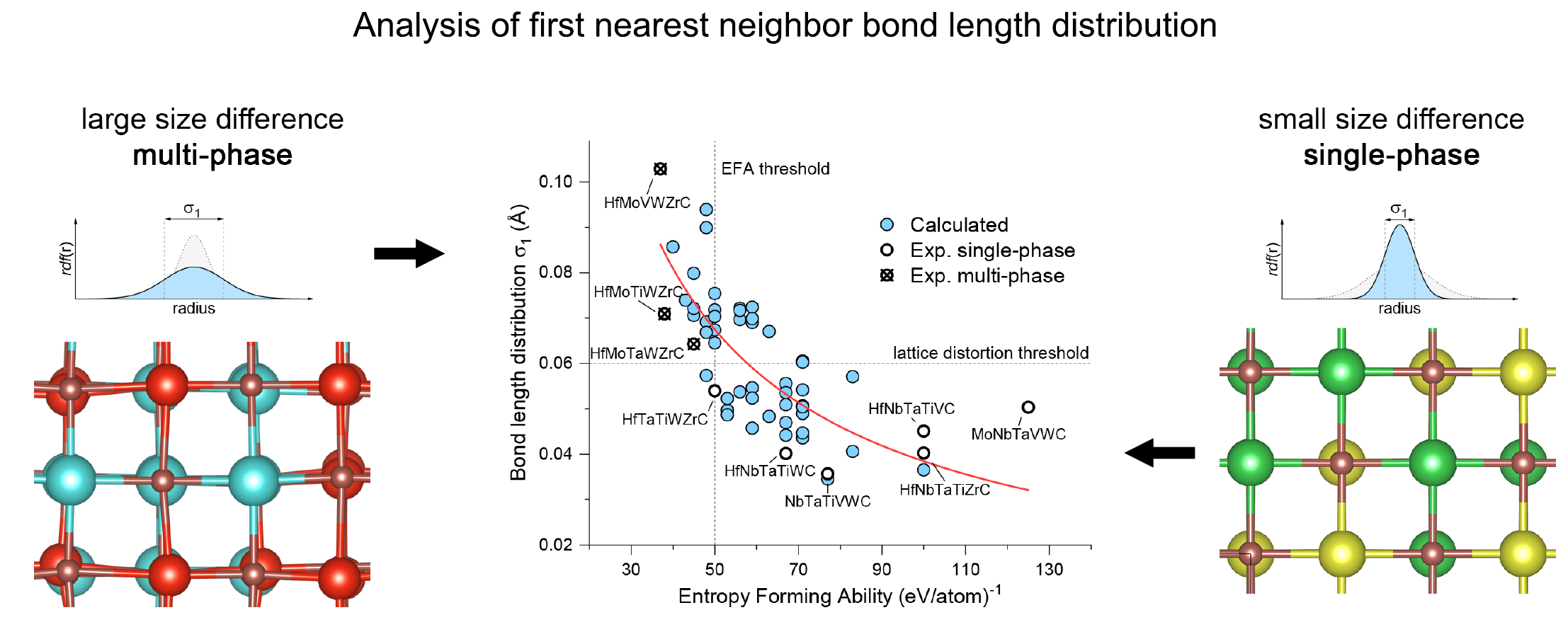}
	}
	\label{fig:Graphical_abstract}
\end{figure}

Keywords: High-entropy, Carbides, Hume-Rothery rules, Solid solutions
\vspace{5mm}

\begin{mainmatter}
\chapter{Introduction}
The discovery of stable ceramic high-entropy materials is a daunting task due to the vast combinatorial space to search in. To speed up the exploration, we rely on descriptors that allow a rough estimate of the chance of success with low effort. One such tool is the Hume-Rothery rules \cite{Hume-Rothery1934}, that provide a simple physical explanation for the formation of solid solutions in metals.

Of these Hume-Rothery rules, the size difference between constituting atoms, abbreviated with $\delta$, is most often used as descriptor to describe the propensity to form solid solutions in high-entropy materials. For example, \cite{Zhang2018a} used a data-driven approach to determine suitable elements for compositionally complex sulfides. \cite{Jiang2018} and \cite{Spiridigliozzi2021} investigated oxides in perovskite and fluorite structure, respectively. \cite{Tang2021} used  bond the bond strength in addition to the length to predict properties of carbides, nitrides, and carbonitrides, while \cite{Liu2017} formulated a modified $\delta$ parameter which also takes the shear modulus mismatch into account. \cite{Zhou2023} found that the most important features in predicting hardness and Young's modulus in carbides with high-entropy sublattice are the valence electron concentration, the deviation of melting temperatures, and the fraction-weighted mean total energies, while atomic size differences played a minor, but still significant role.

In contrast, the entropy forming ability (EFA) \cite{Sarker2018} was formulated as a new descriptor for entropy stabilized compounds, and is sometimes interpreted as a driving force for local ordering, where a sufficiently large value indicates the propensity to form a single-phase solid solution due to configurational entropy. The formalism does not take atomic sizes into account. Intrigued by this descriptor, we investigated the same carbides as \cite{Sarker2018} with ab initio calculations -- implementing special quasi random structures (SQS) -- to analyze the local lattice distortion in order to develop a better understanding of the EFA.

\chapter{Methods}
We performed all ab initio calculations using the Vienna Ab initio Simulation Package (VASP) \cite{Kresse1996,Kresse1999} with projector-augmented plane wave pseudo-potentials with generalized gradient approximated exchange-correlation functionals \cite{Perdew1996}. To analyze the local distortions in the structures, we used $2\times2\times2$ SQS supercells \cite{Wei1990,Gehringer2023}, adapting our method established in \cite{Kretschmer2022} for nitrides to the relevant 5-metal carbides in the phase space of Hf, Mo, Nb, Ta, Ti, V, W, and Zr. To improve the statistical ensemble, we have calculated 10 independent 64-atom cells per composition and use average values over these 10 cells. k-meshes \cite{Pack1977} and other details of the calculations can be found in the supplementary material. We quantify the local lattice distortion with a radial distribution function, in which the parameter $\sigma_1$ describes the distribution of bond lengths in the first coordination sphere (nearest neighbors),

\begin{equation}
	rdf(r)=a_i\cdot e^{-\frac{(r-r_{0,i})^2}{2\sigma_i^2}}\textnormal{,}
\end{equation}

with $r_{0,i}$ as the mean bond length of the $i$th coordination sphere, $a_i$ as a fitting parameter for the distribution peak height, and $\sigma_i^2$ as the variance of the $i$th-neighbor bond length distribution. We contrast this parameter with the commonly used nominal deviation of the average atomic radius $\delta$,

\begin{equation}
	\delta=\sqrt{\sum_{i=1}^{N}X_i\left(1-\frac{r_i}{\overline{r}}\right)^2}\textnormal{,}
\end{equation}

where $X_i$ is the mole fraction of the $i$th component, $r_i$ the nearest neighbor N-metal bond length of the $i$th metal, and $\overline{r}$ the average nearest neighbor N-metal bond length of all metals present \cite{Zhang2008}. Since the actual atomic radii in ceramics can deviate from the radii tabulated for metals, we use the metal-carbon bond lengths of the respective binary face-centered cubic (fcc, NaCl protopye) cells instead of tabulated values to calculate $\delta$.

To complement and confirm our results, we have run more SQS calculations on multicomponent carbides in the same phase space with 2, 3, and 4 metals, and compared the resulting lattice distortion with EFA calculations, which we have prepared with the AFLOW-POCC \cite{Yang2016} module, resulting in 2, 3, and 18 individual structures for carbides with 2, 3, and 4 metals, respectively. Structure files and parameters for all calculations are listed in the supplementary materials.

\chapter{Results and Discussion}

\begin{figure}
	%	\captionsetup{width=19cm}
	\makebox[\textwidth][c]{
		\includegraphics[width=19cm]{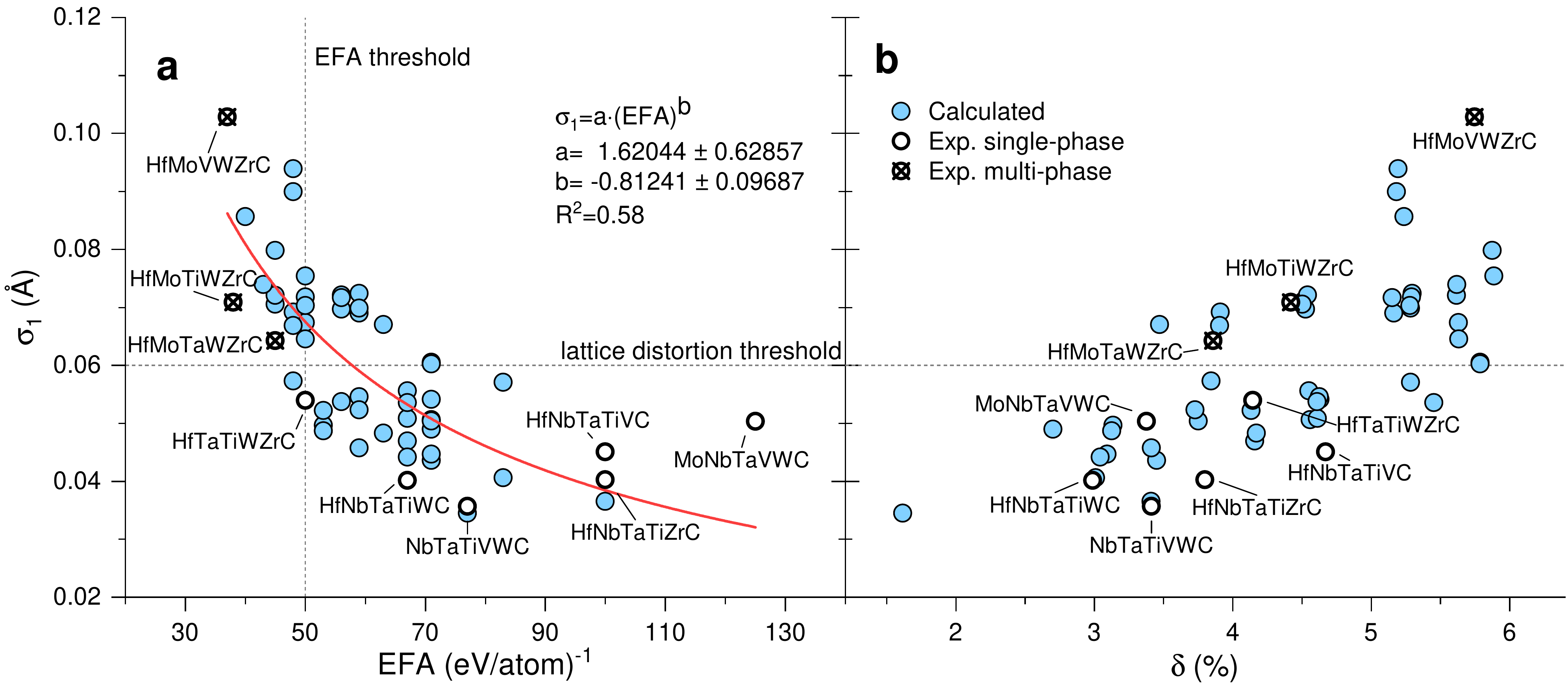}	
	}
	\caption{\textbf{a} The EFA shows a clear correlation with the local lattice distortion, quantified by the distribution of nearest neighbor bond lengths, $\sigma_1$. A power law fit (red line) illustrates the trend. The EFA data and experimental results (bold circles) are taken from \cite{Sarker2018}. \textbf{b} The nominal size variation $\delta$ correlates with $\sigma_1$, but is not precise enough to separate single- and multi-phase compositions its own.}
	\label{fig:EFA_1}
\end{figure}

We have mapped the correlation between $\sigma_1$ and the EFA of the 56 compounds studied in \cite{Sarker2018} in \autoref{fig:EFA_1}\textbf{a}. The threshold for single-phase structures is at EFA=\SI{50}{(eV/atom)^{-1}}, which is shown as a dotted line. A power law fit (red line) illustrates that large lattice distortions occur only in compositions with small EFA values. Our data demonstrates that the EFA depends on size effects of the constituting atoms, which has not been recognized before.

Sarker et al. \cite{Sarker2018} have argued that the formation of a single-phase \ce{(MoNbTaVW)C} is counter-intuitive with the argument, that two of the constituting metals, W, and Mo, prefer other structures and stoichiometric ratios. Our structural data helps to understand this oddity, showing that from the point of atomic size mismatch, the only moderate lattice distortion allows the formation of a single-phase material with these metals. This is especially true, if techniques such as physical vapor deposition or spark plasma sintering with their very fast cooling rates are used, allowing for kinetically-limited single-phase stabilization.

There is significant scattering in the data, causing the exemplary fitting function to display an $R^2$ score of only 0.58, but this fit is only intended to show the general trend. One source of this scattering we found to be in the EFA calculation, which is extremely sensitive to the convergence parameters, this will be discussed in more detail below. The other source is our $\sigma_1$ parameter. When the crystalline lattice shows only very little distortion -- meaning the peak of the radial distribution function is very narrow -- the width of the Gaussian fitting of this narrow peak suffers from numerical inaccuracy, but when the distribution becomes broader, the fit becomes well defined. In effect this means that our $\sigma_1$ parameter cannot reliably quantify small differences in lattices with low distortion, but is effective at discerning lattices with little distortion from more distorted ones. Based on the experimental findings from \cite{Sarker2018}, we can define a rough threshold of $\sigma_1\approx\SI{0.06}{\AA}$, above which the lattice is too distorted to maintain a single phase.

A simpler approach to the lattice distortion is the nominal size mismatch $\delta$, which correlates with $\sigma_1$, see \autoref{fig:EFA_1}\textbf{b}. Despite using metal-carbon bond lengths as size parameter, the $\delta$ parameter fails to separate single- and multi-phase compositions. The underlying principle of the Hume-Rothery rules is still valid, but since atomic radii are influenced by local charge transfer \cite{Casillas-Trujillo2021}, atomic radii are ill defined in complex solid solutions. Therefore, the simple $\delta$ parameter fails to capture the chemical complexity involved. Instead, the local relaxations in our SQS cells provide a better guide to real distortions of the lattice and can be used to gauge the single-phase stability.

\begin{figure}
	%	\captionsetup{width=19cm}
	\makebox[\textwidth][c]{
		\includegraphics[width=19cm]{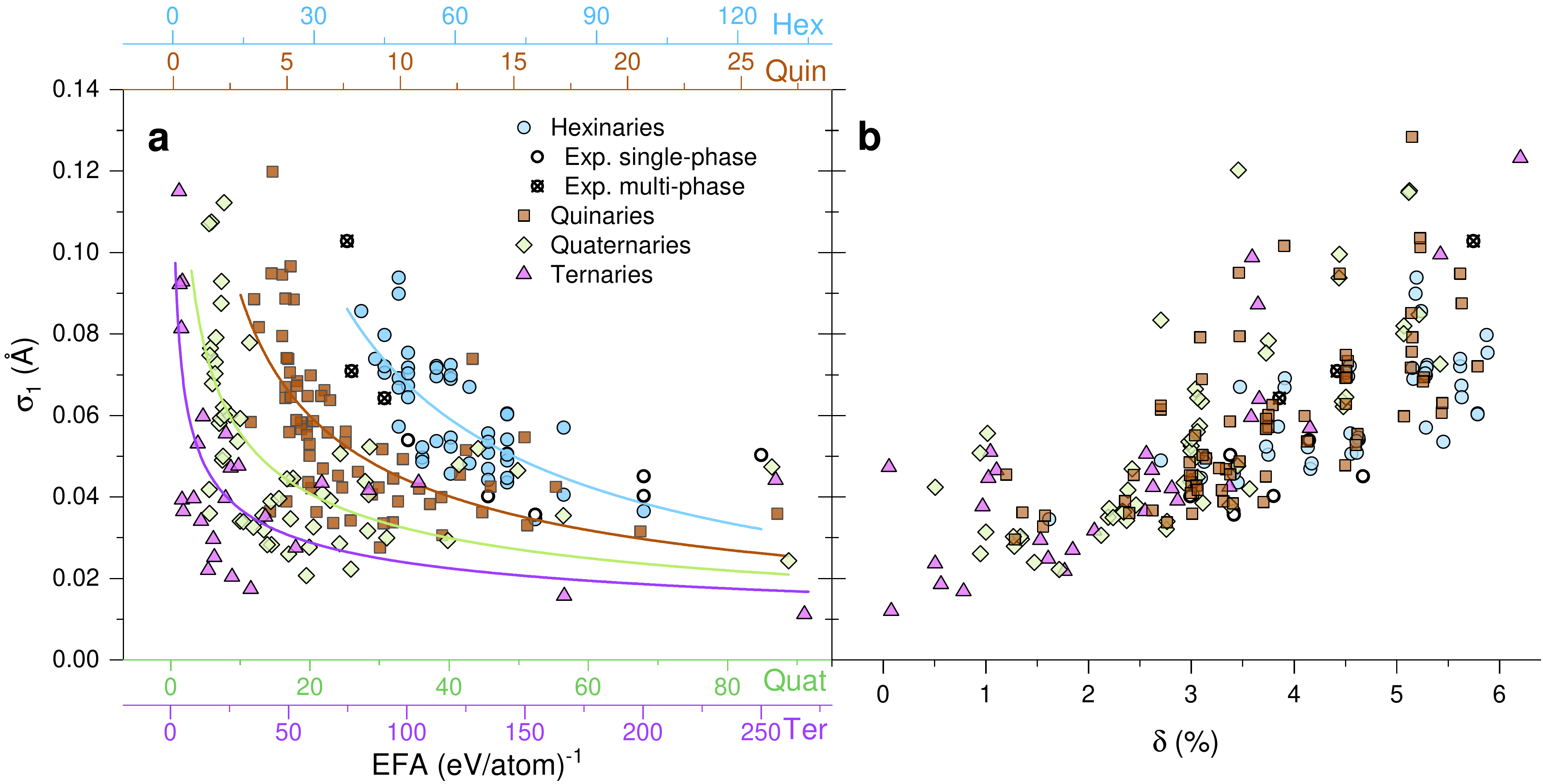}	
	}
	\caption{\textbf{a} The different multinary level alloys all show the same correlation between EFA and $\sigma_1$. The lines represent power law fits  as in \autoref{fig:EFA_1}\textbf{a} for every multinary level. Note, that the scaling of EFA is different for every multinary level. \textbf{b} The correlation betweeen $\delta$ and $\sigma_1$ is also apparent at all multinary levels. Note, that at large values of $\delta$, the lattice distortion $\sigma_1$ is always significant.}
	\label{fig:EFA_2}
\end{figure}

To confirm our finding, we conducted further calculations of the EFA and $\sigma_1$ on the multinary carbides with 2, 3, and 4 of the same metals. The correlation between the additional datasets is depicted in \autoref{fig:EFA_2}\textbf{a} for all multinary levels, demonstrating trends identical to the one shown in \autoref{fig:EFA_1}\textbf{a}. A power law fit for every multinary level is inserted as guide line with the corresponding color. Please note, that the scaling of the EFA is different in this figure for every multinary level. Thus, an EFA of \SI{50}{(eV/atom)^{-1}} is not the general threshold to form a single-phase material.

For our EFA calculations we noticed the extreme sensitivity of the EFA score on the full convergence of the cells. The total energies of the individual relaxed cells differ only by magnitudes in the meV range, so that slightly different -- but high -- stopping criteria for relaxation can lead to wildly different EFA scores. We therefore advise utmost caution when calculating this descriptor.

The relationship between $\sigma_1$ and $\delta$ is shown for all considered compositions in \autoref{fig:EFA_2}\textbf{b}. Large variations in $\sigma_1$ for a given $\delta$ demonstrate only a moderate reliability of the $\delta$ parameter to gauge the actual lattice distortion. But a striking significance of this relationship is the total absence of small lattice distortion in compounds with large differences of atomic radii.

\begin{figure}
	%	\captionsetup{width=19cm}
	\makebox[\textwidth][c]{
		\includegraphics[width=9cm]{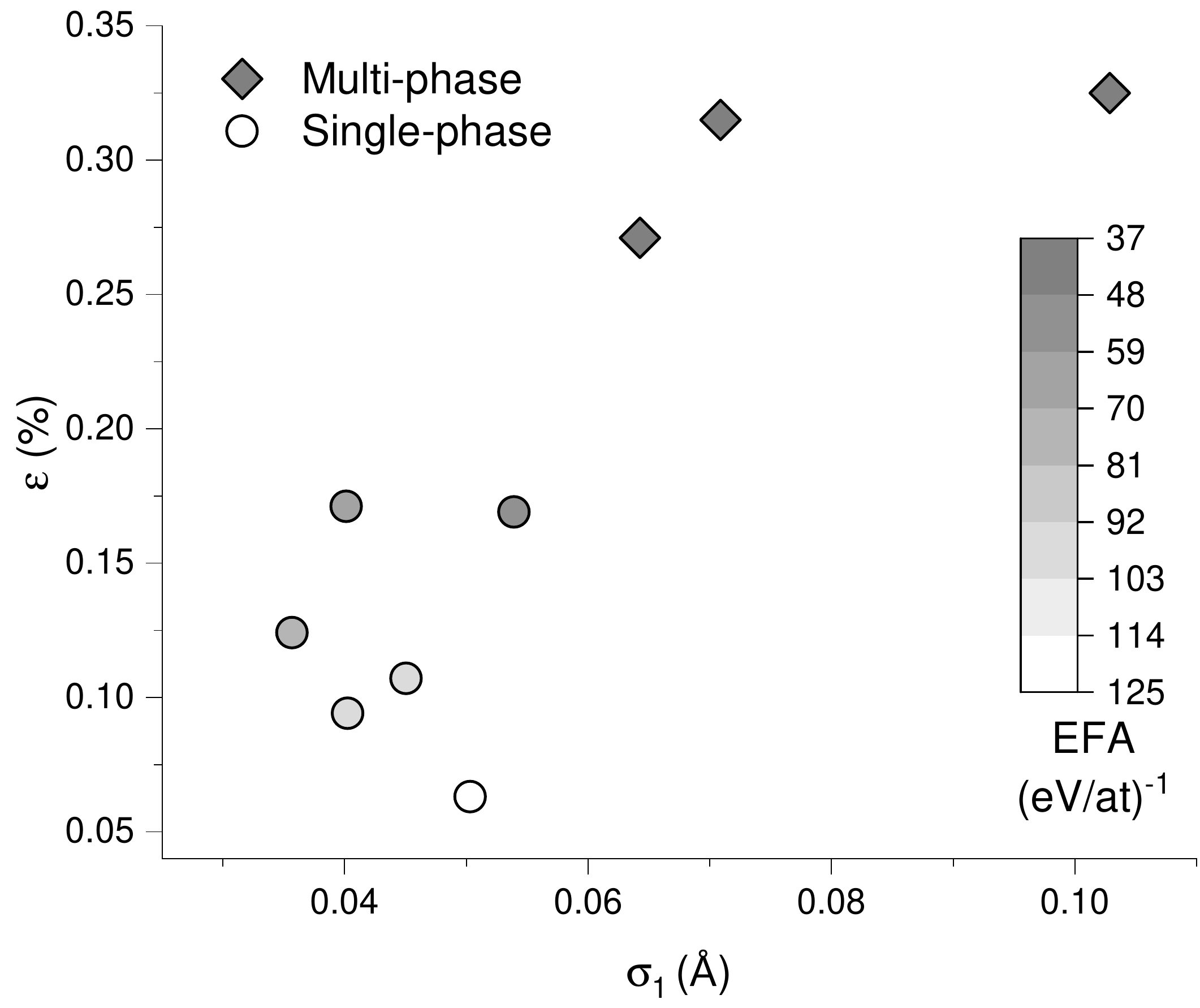}	
	}
	\caption{The calculated lattice distortion $\sigma_1$ correlates well with the XRD measured interplanar spacing $\epsilon$ of the nine experimentally investigated carbides. The grayscale filling denotes the corresponding EFA values. $\epsilon$ and EFA values are taken from \cite{Sarker2018}.}
	\label{fig:EFA_3}
\end{figure}

Additionally, with the different sizes of the constituting atoms -- and thereby roughened crystal planes -- we can easily explain the experimental findings presented in \cite{Sarker2018} over the whole data range. While the interplanar spacing $\epsilon$, taken as the X-ray diffraction-measured reflex width in \cite{Sarker2018}, inversely correlates with the EFA for both single- and multi-phased materials, $\epsilon$ also correlates with the calculated lattice distortion $\sigma_1$, see \autoref{fig:EFA_3}, confirming our calculations.

\chapter{Conclusion}
We show that the EFA descriptor does not screen synthesizability of single-phase materials due to entropy stabilization, but rather atomic size mismatch. The underlying Hume-Rothery rules remain central for the stability of solid solutions even in high-entropy compounds. Our radial distribution analysis of bond lengths in relaxed special quasi-random-structure cells provides a quick and computationally cheap method to gauge the lattice distortion and thus single-phase synthesizability in complex compounds.

\end{mainmatter}

\begin{backmatter}

\section{Acknowledgements}
This work was funded by the Austrian COMET Program (project K2 InTribology, no. 872176). Computational results have been achieved with the Vienna Scientific Cluster. The authors acknowledge TU Wien Bibliothek for financial support through its Open Access Funding Program and thank David Holec (Montanuniversität Leoben) for valuable discussion.

\section{Author contributions}
A.K. conceptualized the work and performed the calculations and analysis. P.H.M. provided computational resources and funding and assisted in the analysis. Both authors contributed to writing the paper.

\section{Data availability}
A part of the raw data can be found in the supplementary material. More data supporting the findings of this study is available from the corresponding author upon request.

\section{Declaration of Interests}
The authors declare that they have no competing interests that could appear to have influenced the content of this paper.
\newpage

\bibliography{EFA.bib}

\end{backmatter}
\end{document}